\documentclass{ws-ijmpd}
\usepackage[super,compress]{cite}

\newcommand{\be}{\begin{equation}}
\newcommand{\ee}{\end{equation}}

\begin{document}
\markboth{G.S. Bisnovatyi-Kogan, M. Merafina}
{Orbital precession and other properties of two-body motion in the presence of dark energy}

%
\catchline{}{}{}{}{}
%

\title{Orbital precession and other properties of two-body motion \\ in the presence of dark energy}

\author{G.S. Bisnovatyi-Kogan}
\address{Space Research Institute, Profsoyusnaya 84/32, Moscow, Russia 117997.\\
National Research Nuclear University MEPhI, Kashira Highway, 31, Moscow, Russia 115409.\\
Moscow Institute of Physics and Technology MIPT,\\
Institutskiy Pereulok, 9, Dolgoprudny, Moscow region, Russia 141701.\\
Department of Physics, University of Rome ``La Sapienza", Rome, Italy.}

\author{M. Merafina}
\address{Department of Physics, University of Rome ``La Sapienza", Rome, Italy}

\maketitle

\begin{history}
\received{Day Month Year}
\revised{Day Month Year}
\end{history}

\begin{abstract}
We consider the Kepler two-body problem in the presence of a cosmological constant $\Lambda$. Several dimensionless parameters characterizing the possible orbit typologies are used to identify open and closed trajectories. The qualitative picture of the two-body motion is described and critical parameters of the problem are found.
\end{abstract}


\ccode{PACS numbers: 95.36.+x; 98.80.Es; 98.65.At}




\section{Introduction}

The discovery of dark energy (DE) in the universe is based on observations of the supernova SN Ia at redshift $z\le 1$\cite{1998Riess,1999Perlmutter}\,, and on measurements of the spectrum of fluctuations of the cosmic microwave background radiation (CMB)\cite{2003Spergel,2004Tegmark}\,. These measurements give a value of the cosmological constant $\Lambda\approx 10^{-56}$ cm$^{-2}$. For review articles considering the role of DE in cosmology, see e.g. Refs.~\refcite{w89,cpt92,c01,r01,oonm18,n18,i15,ds16}.

Before the observational measurements of the DE density in SN Ia and CMB investigations, the limits on the present value of the cosmological constant $\Lambda$ which determined the DE density to be $\rho_{DE}=\Lambda\,c^2/8\pi G$ were obtained using precision measurements of the binary pulsar timing and planetary motion in the Solar System. From measurements of the perihelion shift of Mercury \cite{i83} an upper limit $\Lambda \,< \, 10^{-42}$ cm$^{-2}$ was obtained that decreased  to $\Lambda \,< \, 10^{-55}$ cm$^{-2}$ 15 years later \cite{ct98}. This result was later invalidated in Ref.~\refcite{i06}, where the upper limit  $\Lambda \,< \, 4\cdot 10^{-45}$ cm$^{-2}$ was given, see also Refs.~\refcite{i08,a13,os16,i18}. A theoretical analysis of the influence of the cosmological constant on the gravitomagnetic clock effect and the gravitational time delay of electromagnetic waves, as well as the effect of $\Lambda$ on the pericenter precession, had been done in Ref.~\refcite{khm03}.  Ref.~\refcite{kw03} showed that the measured value of $\Lambda$ has a negligible effect  on the measurement of the perihelion shift of Mercury.

Limits on the cosmological constant from effects on Solar and stellar systems as well as on binary pulsars  have been discussed in Refs.~\refcite{js06,sj06}, where it was claimed that the best constraint comes from the perihelion precession of Earth and Mars, $\Lambda  < 1\cdot 10^{-46}$ cm$^{-2}$. Various Solar system effects in the Schwarzschild-de Sitter space-time had been calculated in Ref.~\refcite{kkl06}. The behavior of Keplerian orbits due to central-force perturbations and cosmological expansion was studied in Refs.~\refcite{amf07,am07,sj07,cs08}. Anthropic constraints on the cosmological constant from the Sun's motion through the Milky Way were discussed in Ref.~\refcite{i10}.

Refs.~\refcite{2001Chernin,2008Chernin} showed that the outer parts of galaxy clusters (GC) may be strongly influenced by dark energy. If we consider the relative motion of two rich clusters, then we should deal with sizes for which the influence of the DE in the form of the cosmological constant $\Lambda$ is important. We consider here a simplified problem of the relative motion of two rich clusters represented by two point masses.

The analytic solution of the problem of two-body motion in the presence of a nonzero $\Lambda$ in the quasi-Newtonian approximation was given in Ref.~\refcite{2013Emelyanov}, using the table of integrals from Ref.~\refcite{ryzhik}. Numerical approximations were used to calculate the elliptical integrals used in this analytic solution.
We solve this problem using a different mathematical approach and identify the main critical parameters for the two-body system, which had not been calculated in that article. Our calculations describe orbital precession for non-circular motion in the presence of DE, and evaluate the associated periods and quasi-periods in terms of various parameters of the system.

\section{Equations for two-body problem in the presence of $\Lambda$ in the quasi-Newtonian approximation}

We use the equations governing Keplerian motion in the presence of nonzero $\Lambda$ given in Ref.~\refcite{bkm2019}. For two masses $m_1$ and $m_2$ rotating around each other, these equations
describe the behaviour of their separation vector in the plane of the motion characterized by its length $r$ and polar angle $\varphi$ in a polar coordinate system as in the Kepler problem in Newtonian gravity \cite{1969Landau}. Introducing the reduced mass $\mu$, the total mass $M$, and the conserved value of the angular momentum $L$ by
\begin{equation}
\mu=m_1m_2/(m_1+m_2), \quad M=m_1+m_2, \quad L=\mu r^2\dot{\varphi} ,
\label{eq1}
\end{equation}
the equations of motion take the form \cite{bkm2019}
\begin{equation}
{\ddot r} = \,-\frac{GM}{r^2} +\frac{L^2}{\mu^2 r^3}+\frac{\Lambda c^2}{3} r, \qquad
\dot{\varphi}=\frac{L}{\mu r^2}.
\label{eq2}
\end{equation}
Integrating the first equation in (\ref{eq2}), we obtain the expression for the conserved total energy $E$ of the system in the form
\begin{equation}
E=\frac{1}{2}\mu\dot{r}^2+\mu\left(-\frac{GM}{r}+\frac{L^2}{2\mu^2 r^2}-\frac{\Lambda c^2}{6}r^2\right).
\label{eq3}
\end{equation}

There are two characteristic radii in this problem: $r_0$, at which gravity of the binary system is balanced by the antigravity of DE (zero-gravity radius) \cite{2008Chernin}, and the Keplerian radius  $r_k$ of the circular orbit in Newtonian gravity \cite{1969Landau}, namely
\begin{equation}
r_0=\left(\frac{3GM}{\Lambda c^2}\right)^{1/3}\ , \qquad   r_k=\frac{L^2}{GM\mu^2} \ .
\label{eq4}
\end{equation}
The equation for $\dot r$ following from (\ref{eq3}) is
\begin{equation}
\dot r =\pm\,\sqrt{2\frac{E}{\mu}-2\left(-\frac{GM}{r}+\frac{L^2}{2\mu^2 r^2}-\frac{\Lambda c^2}{6}r^2\right)}\,,
\label{eq5}
\end{equation}
where the ``$\pm$" sign distinguishes the increasing and decreasing phases of the radial motion. From (\ref{eq5})  and the second equation in (\ref{eq2}), we obtain the relation connecting $r$ and $\varphi$ along the trajectory
\begin{equation}
\frac{dr}{d\varphi}=\pm\,r\,\sqrt{2\frac{E\mu}{L^2}r^2+2\frac{GM\mu^2}{L^2}r-1+\frac{\Lambda c^2\mu^2}{3L^2}r^4}\ .
\label{eq6}
\end{equation}

\subsection{Dimensionless variables}

Introducing the dimensionless radius $x$ and dimensionless time $\tau$, the second 
equation in (\ref{eq2}) becomes
\begin{equation}
x=\frac{r}{r_k}\ ,\qquad \tau=\frac{L}{\mu r_k^2}t\ ,\qquad \frac{d\varphi}{d\tau}=\frac{1}{x^2}\ .
\label{eq7}
\end{equation}
Rewriting equations (\ref{eq5}) and (\ref{eq6}), taking into account (\ref{eq4}) and (\ref{eq7}), leads to
\begin{equation}
\frac{dx}{d\tau} =\pm\,\frac{1}{x}\,\sqrt{2\frac{E r_k}{GM\mu}x^2+2x
-1+\left(\frac{r_k}{r_0}\right)^3 x^4}
\label{eq8}
\end{equation}
and
\begin{equation}
\frac{dx}{d\varphi} =\pm\,x\,\sqrt{2\frac{E r_k}{GM\mu}x^2+2x
-1+\left(\frac{r_k}{r_0}\right)^3 x^4}\ .
 \label{eq9}
\end{equation}

As in the Kepler problem it is convenient to use the reciprocal variable $u=1/x$ instead of $x$ and introduce the dimensionless parameter $d=(r_k/r_0)^3$. From equations (\ref{eq8}) and (\ref{eq9}), we obtain (the upper sign corresponds to $\dot{r}>0$)
\begin{equation}
\frac{d\varphi}{d\tau}=u^2,\qquad \frac{du}{d\tau}=\mp u^2\sqrt{-u^2+2u+\varepsilon +d/u^2}
\label{eq10}
\end{equation}
and then
\begin{equation}
\frac{du}{d\varphi} =\mp\sqrt{-u^2+2u+\varepsilon +d/u^2}\ ,
\label{eq11}
\end{equation}
where $\varepsilon=2Er_k/GM\mu$ and $d=\Lambda c^2r_k^3/3GM$.

\section{Keplerian limit}

In the absence of DE ($\Lambda =0$ implies $d=0$), equations (\ref{eq10}) and (\ref{eq11}) reduce to
\begin{equation}
\frac{du}{d\tau}=\mp u^2\sqrt{-u^2+2u+\varepsilon}\ ,\qquad \frac{du}{d\varphi}=\mp\sqrt{-u^2+2u+\varepsilon}\ ,
\label{eq12}
\end{equation}
where $-1\le\varepsilon<0$ for closed orbits. Consider the branch with the sign ``+", writing the solution in the form
\begin{equation}
\Delta\tau =\int_{u_-}^{u_+}\frac{du}{u^2\sqrt{-u^2+2u+\varepsilon}}, \qquad
\Delta\varphi =\int_{u_-}^{u_+}\frac{du}{\sqrt{-u^2+2u+\varepsilon}}\ .
\label{eq13}
\end{equation}
Here the roots $u_+= 1+\sqrt{1+\varepsilon}$ and $u_-= 1-\sqrt{1+\varepsilon}$ of the expression inside the square root are related to the minimal (pericenter) and maximal (apocenter) separation between the two bodies, respectively. Integration leads to
\begin{equation}
\Delta\varphi =\int_{1-\sqrt{1+\varepsilon}}^{1+\sqrt{1+\varepsilon}}
\frac{du}{\sqrt{1+\varepsilon-(u-1)^2}}=\int_{-\sqrt{1+\varepsilon}}^{+\sqrt{1+\varepsilon}}
\frac{dv}{\sqrt{1+\varepsilon-v^2}}=\int_{-1}^{+1}\frac{dz}{\sqrt{1-z^2}}=\pi\ .
\label{eq14}
\end{equation}
The first integral in (\ref{eq13}) can be evaluated analytically \cite{ryzhik}. We have then
\begin{equation}
\Delta\tau = \left[{-\frac{\sqrt{-u^2+2u+\varepsilon}}{\varepsilon u}+\frac{1}{(-\varepsilon)^{3/2}} \arcsin\frac{u+\varepsilon}{u\sqrt{1+\varepsilon}}}\right]_{u_-}^{u_+} =\frac{\pi}{(-\varepsilon)^{3/2}}\ .
\label{eq15}
\end{equation}
The results obtained in (\ref{eq14}) and (\ref{eq15}) are related to half of the periodic trajectory, which is closed for Keplerian motion, with the change of the angle equal to $2\pi$ during a cycle. The ``-" sign in (\ref{eq12}) is related to the second half of the closed elliptical trajectory, describing the motion from pericenter to apocenter, with a decreasing $u$-velocity ($\dot{u}<0$, corresponding to $\dot{r}>0$).

\begin{figure}[t]
\includegraphics[width=1.0\textwidth]{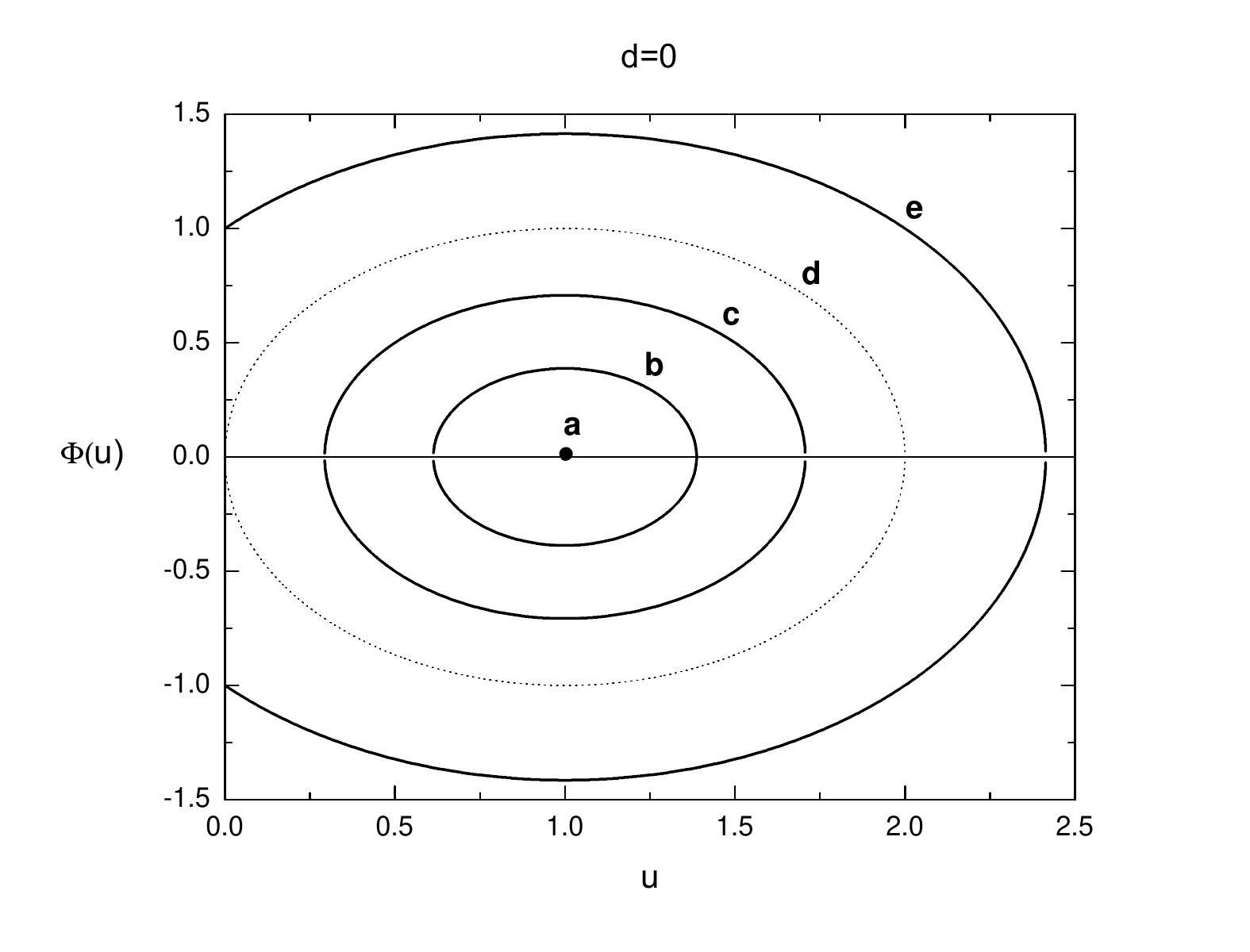}
\caption{The family of Keplerian orbits described by the quantity $\Phi(u)=\mp\sqrt{-u^2+2u+\varepsilon}$ as a function of $u=r_k/r$ at fixed values of $d=0$, for various values of the dimensionless total energy of the system 
$\varepsilon=2Er_k/GM\mu$ (see also Table I). The curves correspond to the
values a) $\varepsilon=-1$; b) $\varepsilon=-0.85$; c) 
$\varepsilon=-0.5$; d) $\varepsilon=0$; e) $\varepsilon=1$. The zeros of $\Phi$ (roots) are the turning points of the trajectory: left 
$(u_-)$ and right $(u_+)$ zeros correspond to the apocenter and pericenter of the trajectory, respectively. Circular orbit (a) corresponds to the black spot at $u=1$ and $\Phi=0$. Dotted curve (d) with a root at $u=0$ corresponds to the parabolic motion 
$(\varepsilon=0)$. Curves with $\varepsilon >0$ correspond to unbound systems (hyperbolic trajectories). No solutions exist for 
$\varepsilon<-1$.
\label{fig1}}
\end{figure}

The dimensional period of the Keplerian motion $P_k$, taking into account equations (\ref{eq11}) and (\ref{eq15}), can be rewritten as (see also Ref.~\refcite{1969Landau})
\begin{equation}
P_k= 2 \Delta t =\frac{2\pi}{(-\varepsilon)^{3/2}}\frac{\mu r_k^2}{L}=\pi GM\left[\frac{\mu^3}{2(-E)^3}\right]^{1/2}\ ,\quad {\rm since}\ \ \Delta t=\frac{\mu r_k^2}{L}\Delta\tau\ .
\label{eq16}
\end{equation}
The trajectory of the Keplerian motion is obtained from the indefinite integral
\begin{equation}
\varphi =\int\frac{du}{\sqrt{-u^2+2u+\varepsilon}}=\arcsin\frac{u-1}{\sqrt{1+\varepsilon}} + {\rm const}\ .
\label{eq17}
\end{equation}
Choosing const\,=\,$\pi/2$, which corresponds to $\varphi=0$ at the apocenter of the trajectory, and returning to the dimensional variables, we obtain finally
\begin{equation}
r=\frac{r_k}{1+\sqrt{1+\varepsilon}\sin(\varphi-\pi/2)}=\frac{r_k}{1+e\sin(\varphi-\pi/2)}\ .
\label{eq18}
\end{equation}
Here, the quantity $e=\sqrt{1+\varepsilon}$ is the eccentricity of the elliptical trajectory  in the Keplerian motion described by the equation
(\ref{eq18}). The whole family of Keplerian trajectories in the $(u,\Phi)$ plane
is plotted in Fig.~\ref{fig1}. The quantity $\Phi(u)=du/d\varphi=\mp\sqrt{-u^2+2u+\varepsilon}$ is the angular velocity of the $u$-variable. The main parameters of the Keplerian trajectories ($d=0$) of Fig.~\,\ref{fig1} are also given in Table\,I.

\begin{table}[h]
{{\bf Table I.} Trajectory parameters of two body motion for $d=0$ (Keplerian orbits, without DE). The value $u_M$ corresponds to the maximum of the positive branch of each curve. Letters in the ``trajectory type" column refer to the labels in Fig.~\,\ref{fig1}.}

\medskip
\centering
\scriptsize
\begin{tabular}{ccccl}
\hline
$\varepsilon$ & $u_-$ & $u_+$ & $u_M$ & trajectory type\\
\hline
\\
-1   	&1			&1		&1	&a) circular (black spot) \\
-0.85	&0.61270	&1.3873	&1	&b) elliptical \\
-0.5	&0.29289	&1.7071	&1	&c) elliptical \\
0		&0			&2		&1	&d) parabolic (dotted) \\
1		&-			&2.4142	&1	&e) hyperbolic \\
\\
\hline
\end{tabular}
\end{table}

\section{Trajectories of the two body motion in the presence of DE}

\begin{figure}[t]
\includegraphics[width=1.0\textwidth]{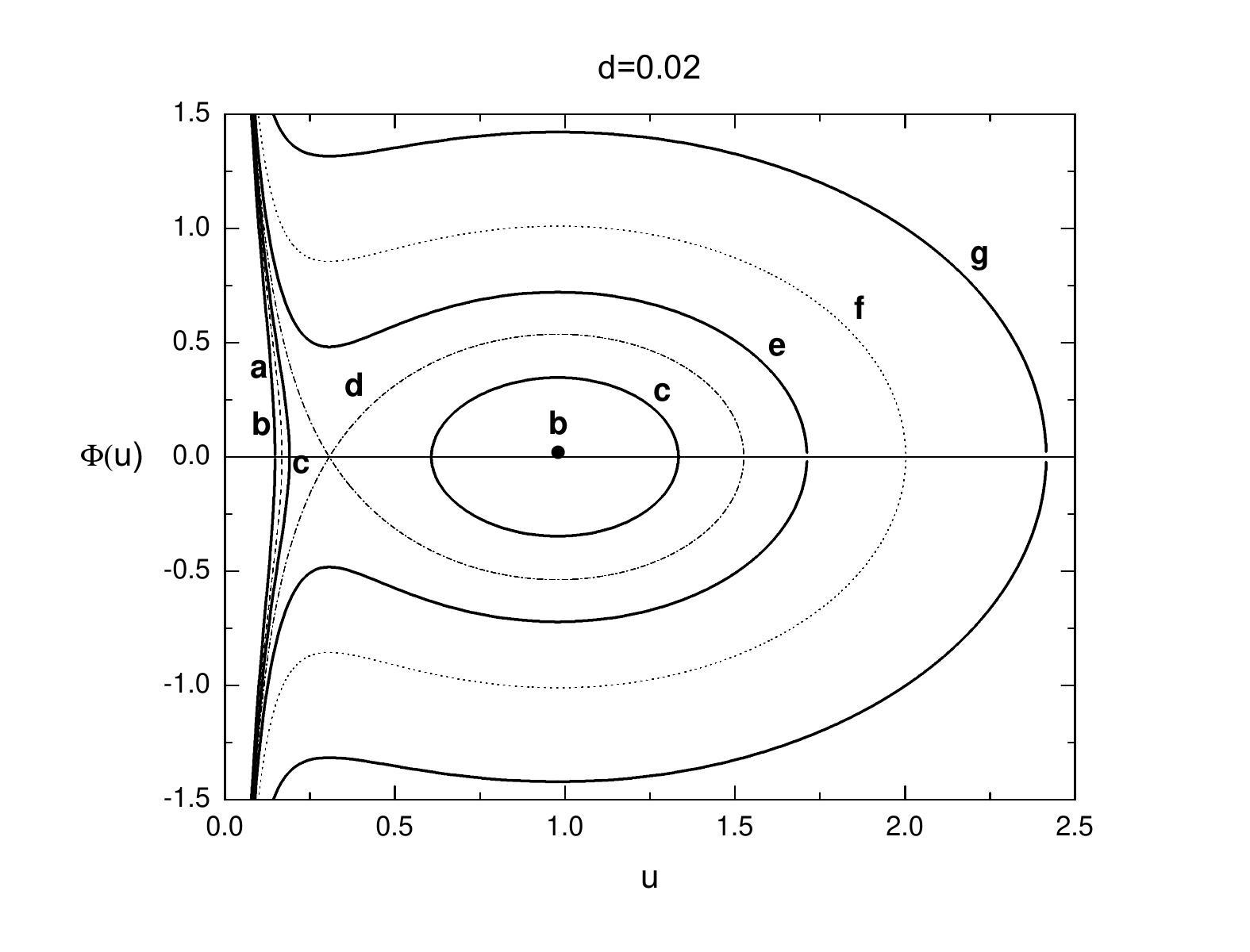}
\caption{The family of trajectories described by the quantity $\Phi(u)=\pm\sqrt{-u^2+2u+\varepsilon+d/u^2}$ as a function of $u=r_k/r$ at the fixed value $d=0.02$, for selected values of the dimensionless total energy of the system 
$\varepsilon=2Er_k/GM\mu$ (see also Table II). The curves correspond to the
values a) $\varepsilon=-1.2$; b) $\varepsilon=-1.02043$; c) $\varepsilon=-0.9$; d) $\varepsilon=-0.731954$; e) $\varepsilon=-0.5$; f) $\varepsilon=0$; g) $\varepsilon=1$. For each value of $\varepsilon$ in the range $-1.02043<\varepsilon<-0.731954$ we have one hyperbolic (unbound) and one quasi-elliptical (bound) trajectory. The zeros of $\Phi$ for bound orbits are the turning points of the trajectory: left $(u_-)$ and right $(u_+)$ zeros correspond to the apocenter and pericenter of the trajectory, respectively. The circular orbit (b) corresponds to the black spot at $u=0.978663$ and $\Phi=0$. The crossing in the dash-dotted curve (d) at $u=0.306691$ and $\Phi=0$ identifies an instability point on the trajectory (transition). Zeros of $\Phi$ in unbound (infinite) curves refer to the pericenter of pure hyperbolic $(u_0)$ or semi-hyperbolic $(u_+)$ trajectories. Finally, the dotted curve (f) corresponds to zero system total energy $(\varepsilon=0)$.
\label{fig2}}
\end{figure}

\begin{figure}[t]
\includegraphics[width=1.0\textwidth]{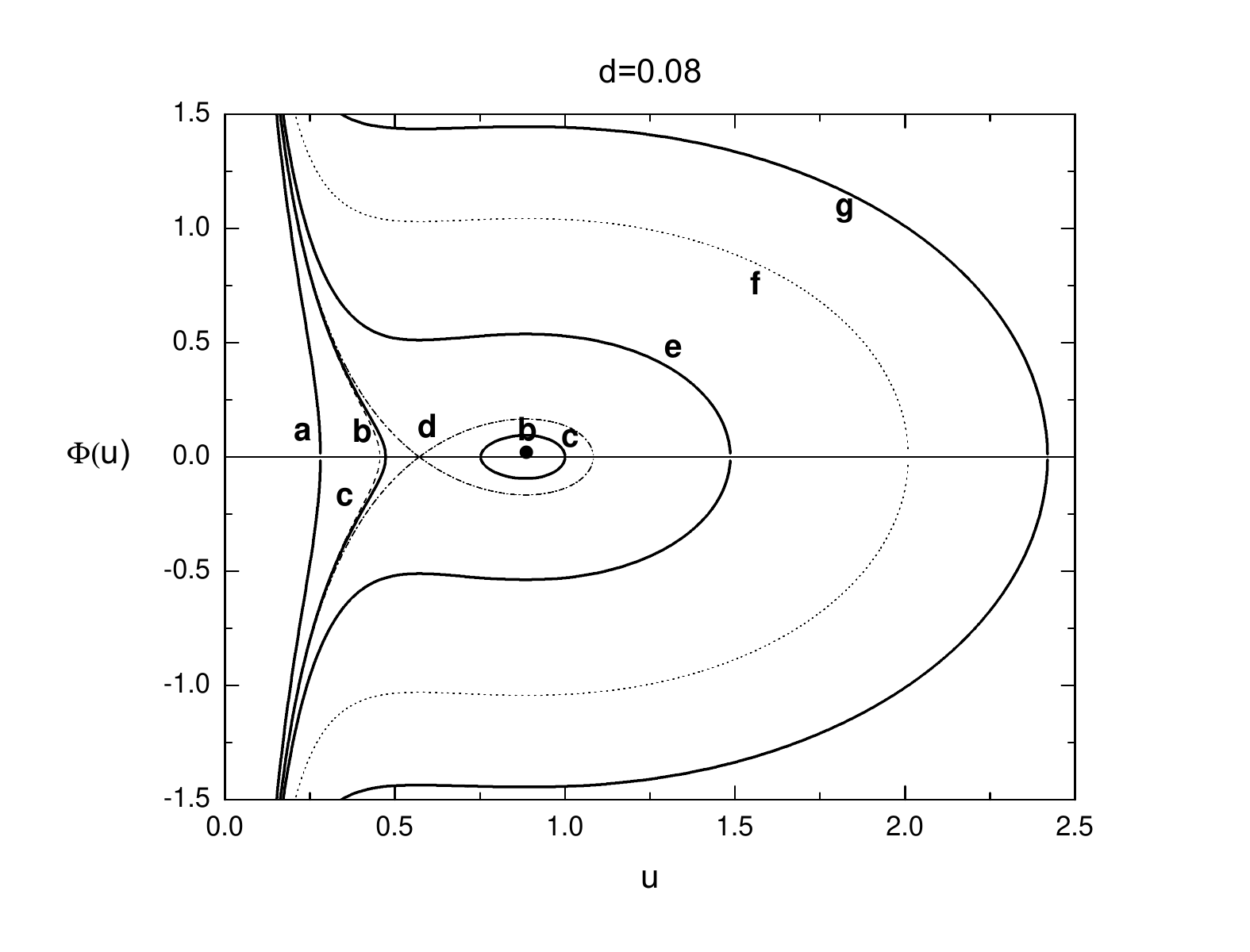}
\caption{The family of trajectories described by the quantity $\Phi(u)=\pm\sqrt{-u^2+2u+\varepsilon+d/u^2}$ as a function of $u=r_k/r$ at fixed value of $d=0.08$, for selected values of the dimensionless total energy of the system 
$\varepsilon=2Er_k/GM\mu$ (see also Table III). The curves correspond to the values a) $\varepsilon=-1.5$; b) $\varepsilon=-1.08892$; c) $\varepsilon=-1.08$; d) $\varepsilon=-1.06133$; e) $\varepsilon=-0.8$; f) $\varepsilon=0$; g) $\varepsilon=1$. For each single value of $\varepsilon$ in the range $-1.08892<\varepsilon<-1.06133$ we have one hyperbolic (unbound) and one quasi-elliptical (bound) trajectory. Zeros of $\Phi$ (roots) in bound orbits are the turning points of the trajectory: left $(u_-)$ and right $(u_+)$ zeros correspond to the apocenter and pericenter of the trajectory, respectively. The circular orbit (b) corresponds to the black spot at $u=0.884319$ and $\Phi=0$. The crossing in the dash-dotted curve (d) at $u=0.571571$ and $\Phi=0$ identifies an instability point on the trajectory (transition). Zeros of $\Phi$ in unbound (infinite) curves refer to the pericenter of pure hyperbolic $(u_0)$ or semi-hyperbolic $(u_+)$ trajectories. Finally, the dotted curve (f) corresponds to zero system total energy $(\varepsilon=0)$.
\label{fig3}}
\end{figure}

\begin{figure}[t]
\includegraphics[width=1.0\textwidth]{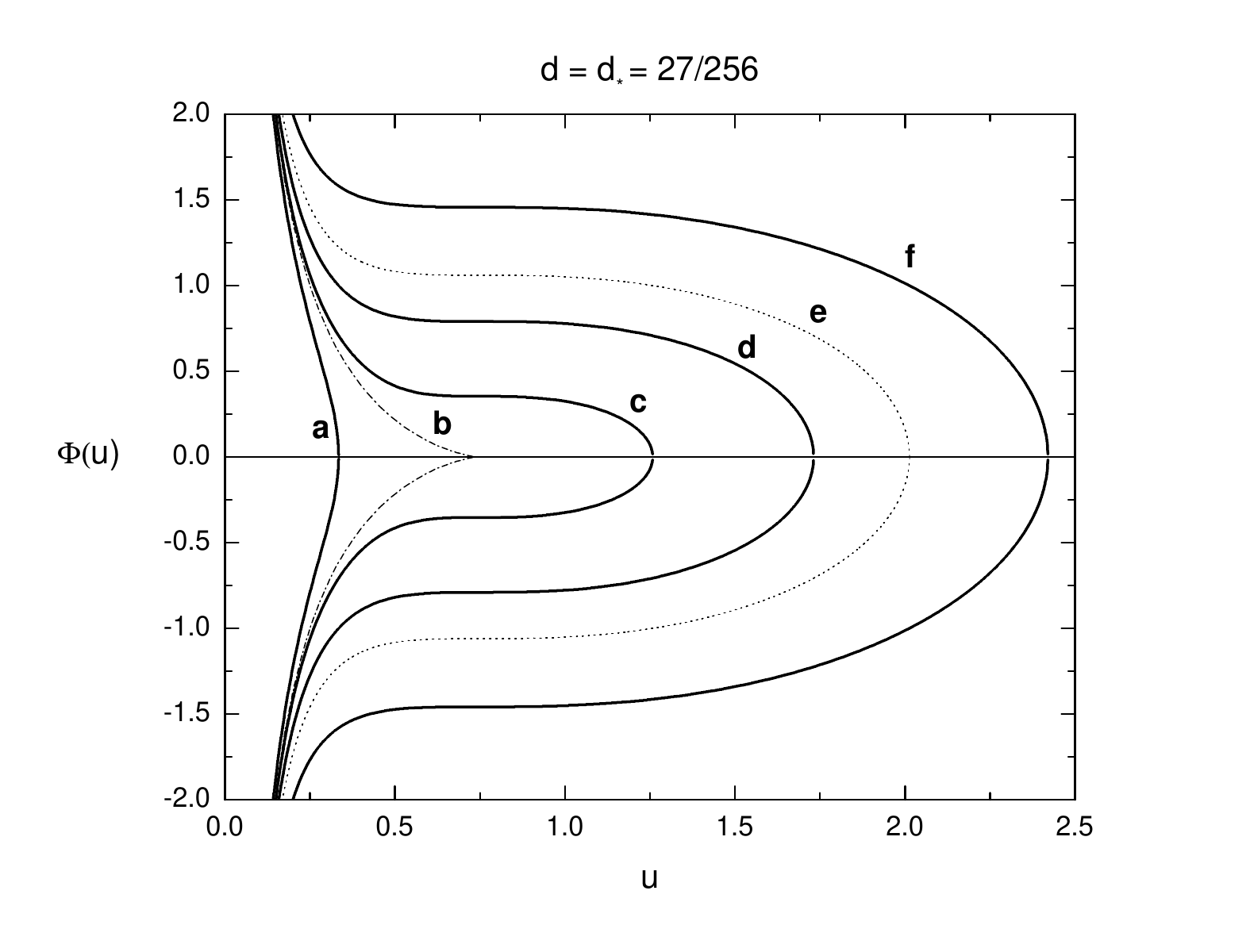}
\caption{The family of trajectories described by the quantity $\Phi(u)=\pm\sqrt{-u^2+2u+\varepsilon+d/u^2}$ as a function of $u=r_k/r$ at fixed value of $d=d_*=27/256$, for selected values of the dimensionless total energy of the system 
$\varepsilon=2Er_k/GM\mu$ (see also Table IV). The curves correspond to the
values a) $\varepsilon=-1.5$; b) $\varepsilon=-9/8$; c) $\varepsilon=-1$; d) $\varepsilon=-0.5$; e) $\varepsilon=0$; f) $\varepsilon=1$. In this case, there are no values of $\varepsilon$ corresponding to bound trajectories. The cusp in the dash-dotted curve (b) at $u=0.75$ and $\Phi=0$ identifies an instability point on the trajectory (transition). Zeros of $\Phi$ of the unbound (infinite) curves refer to the pericenter of pure hyperbolic $(u_0)$ or semi-hyperbolic $(u_+)$ trajectories. Finally, the dotted curve (e) corresponds to zero system total energy $(\varepsilon=0)$.
\label{fig4}}
\end{figure}

\begin{figure}[t]
\includegraphics[width=1.0\textwidth]{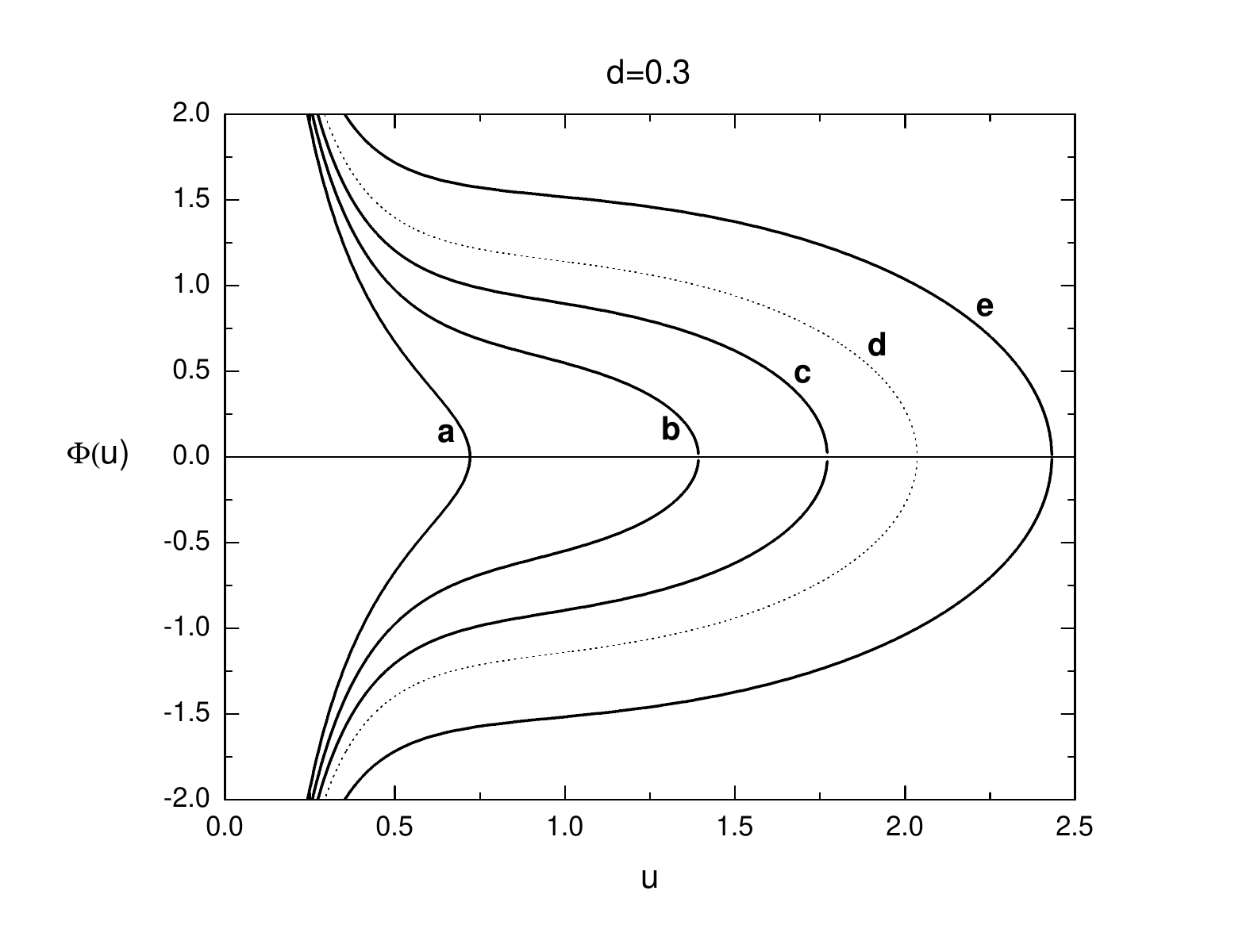}
\caption{The family of trajectories described by the quantity $\Phi(u)=\pm\sqrt{-u^2+2u+\varepsilon+d/u^2}$ as a function of $u=r_k/r$ at fixed value of $d=0.3$, for selected values of the dimensionless total energy of the system 
$\varepsilon=2Er_k/GM\mu$ (see also Table V). The curves correspond to the
values a) $\varepsilon=-1.5$; b) $\varepsilon=-1$; c) $\varepsilon=-0.5$; d) $\varepsilon=0$; e) $\varepsilon=1$. There are no values of $\varepsilon$ corresponding to bound trajectories. Zeros of $\Phi$ of the unbound (infinite) curves refer to the pericenter of semi-hyperbolic $(u_+)$ trajectories. Finally, the dotted curve (d) corresponds to zero system total energy $(\varepsilon=0)$.
\label{fig5}}
\end{figure}

The two-body motion in the presence of DE exhibits qualitatively different behavior with particular features of the motion at certain ranges of the values $d$ and $\varepsilon$. Here $\Phi(u)=\pm\sqrt{-u^2+2u+\varepsilon+d/u^2}$, with $d>0$. The trajectories in the 
$(u,\Phi)$ plane are plotted in Figs.\,\ref{fig2}-\ref{fig5} for selected values of $\varepsilon$ at four fixed values of $d$. Analogously, numerical parameters referring to the figures are given in Tables\,II-V, respectively.

\begin{table}[h]
{{\bf Table II.} Trajectory parameters of the two body motion for $d=0.02$ (presence of DE). The values $u_m$ and $u_M$ correspond to the minimum and maximum of the positive branch of each curve, respectively. Letters in the ``trajectory type" column refer to the labels in Fig.~\,\ref{fig2}.}

\medskip
\centering
\scriptsize
\begin{tabular}{ccccccl}
\hline
$\varepsilon$ & $u_0$ & $u_-$ & $u_+$ & $u_m$ & $u_M$ & trajectory type\\
\hline
\\
-1.2		& 0.14681	& -			& -			& -			& -			& a) infinite \\
-1.02043	& 0.16741	& 0.97866	& 0.97866	& -			& 0.97866	& b) infinite (dashed) + circular (black spot) \\
-0.9		& 0.18951	& 0.60728	& 1.3335	& -			& 0.97866	& c) infinite + quasi-elliptical \\
-0.73195	& 0.30669	& 0.30669	& 1.5260	& 0.30669	& 0.97866	& d) transition-infinite (dash-dotted )\\
-0.5		& -			& -			& 1.7119	& 0.30669	& 0.97866	& e) infinite \\
0			& -			& -			& 2.0025	& 0.30669	& 0.97866	& f) infinite (dotted) \\
1			& -			& -			& 2.4154	& 0.30669	& 0.97866	& g) infinite \\
\\
\hline
\end{tabular}
\end{table}

\begin{table}[b]
{{\bf Table III.} Trajectory parameters of the two body motion for $d=0.08$ (presence of DE). The values $u_m$ and $u_M$ correspond to the minimum and maximum of the positive branch of each curve, respectively. Letters in the ``trajectory type" column refer to the labels in Fig.~\,\ref{fig3}.}

\medskip
\centering
\scriptsize
\begin{tabular}{ccccccl}
\hline
$\varepsilon$ & $u_0$ & $u_-$ & $u_+$ & $u_m$& $u_M$ & trajectory type\\
\hline
\\
-1.5		&0.28034	&-			&-			&-			&-			&a) infinite \\
-1.08892	&0.45580	&0.88432	&0.88432	&-			&0.88432	&b) infinite (dashed) + circular (black spot) \\
-1.08		&0.47267	&0.75231	&1			&-			&0.88432	&c) infinite + quasi-elliptical \\
-1.06133	&0.57157	&0.57157	&1.08296	&0.57157	&0.88432	&d) transition-infinite (dash-dotted) \\
-0.8		&-			&-			&1.4860		&0.57157	&0.88432	&e) infinite \\
0			&-			&-			&2.0099		&0.57157	&0.88432	&f) infinite (dotted) \\
1			&-			&-			&2.4190		&0.57157	&0.88432	&g) infinite \\
\\
\hline
\end{tabular}
\end{table}

In order to analyze the features of the different curves, we calculate the roots of the equation $\Phi(u)=0$ by considering the expression $-u^2\Phi^2(u)=0$. We obtain the following fourth degree equation
\begin{equation}
u^4-2u^3-\varepsilon u^2-d=0\ .
\label{eq19}
\end{equation}
This equation has four roots, but one root always lies in the range $u<0$ so is irrelevant. The other three roots depend on the values of parameters $\epsilon$ and $d$, and sometimes two of them are not real. Therefore we have a maximum of three real roots (or only one) for $\Phi$. These roots are calculated numerically and will be designated by $u_0$, $u_-$ and $u_+$. The analytic solution of the equation $\Phi(u)=0$ was considered in Ref.~\refcite{2013Emelyanov} via elliptical integrals using formulas from Ref.~\refcite{ryzhik}, but this does not make the problem easier because the elliptical integrals must be evaluated numerically.

It is also useful when calculating $\Phi'(u)$ to evaluate the minimum $(u_m)$ and maximum  $(u_M)$ values of $\Phi(u)$. From the positive branch of $\Phi$, we obtain
\begin{equation}
\Phi'(u)=\frac{-u+1-d/u^3}{\sqrt{-u^2+2u+\varepsilon+d/u^2}}\ .
\label{eq20}
\end{equation}
To analyze the extrema we transform the condition $\Phi'(u)=0$ to
\begin{equation}
u^4-u^3+d=0\ .
\label{eq21}
\end{equation}
The solutions of this equation do not depend on $\varepsilon$. The solutions can be calculated graphically by considering the functions $y_1=-d/u^3$ and $y_2=u-1$ and varying the parameter $d$. We have two real solutions (one minimum and one maximum of $\Phi$) for $d\le 27/256$ and no real solutions (no extrema) for $d>27/256$. In particular, for $d=d_*=27/256$ we have a unique solution corresponding to an inflection point at $u=3/4$. For $\varepsilon=-9/8$ the inflection point lies at $\Phi(u)=0$, generating a cusp (see Fig.~\,\ref{fig4}). Moreover, the condition $d>0$ in (\ref{eq21}) implies that $0<u<1$.

\begin{table}[t]
{{\bf Table IV.} Trajectory parameters of the two body motion for $d=d_*=27/256$ (presence of DE). The values $u_m$ and $u_M$ correspond to the minimum and maximum of the positive branch of each curve, respectively. Letters in the ``trajectory type" column refer to the labels in Fig.~\ref{fig4}.}

\medskip
\centering
\scriptsize
\begin{tabular}{ccccccl}
\hline
$\varepsilon$ & $u_0$ & $u_-$ & $u_+$ & $u_m$& $u_M$ & trajectory type\\
\hline
\\
-1.5	&0.33443	&-		&-			&-		&-		&a) infinite \\
-9/8	&0.75		&0.75	&0.75		&0.75	&0.75	&b) transition-infinite (dash-dotted) \\
-1		&-			&-		&1.2581		&0.75	&0.75	&c) infinite \\
-0.5	&-			&-		&1.7316		&0.75	&0.75	&d) infinite \\
0		&-			&-		&2.0129		&0.75	&0.75	&e) infinite (dotted) \\
1		&-			&-		&2.4206		&0.75	&0.75	&f) infinite \\
\\
\hline
\end{tabular}
\end{table}

\begin{table}[t]
{{\bf Table V.} Trajectory parameters of the two body motion for $d=0.3$ (presence of DE). Letters in the ``trajectory type" column refer to the labels in Fig.~\ref{fig5}.}

\medskip
\centering
\scriptsize
\begin{tabular}{ccccccl}
\hline
$\varepsilon$ & $u_0$ & $u_-$ & $u_+$ & $u_m$& $u_M$ & trajectory type\\
\hline
\\
-1.5	&-	&-	&0.72029	&-	&-	&a) infinite \\
-1		&-	&-	&1.3932		&-	&-	&b) infinite \\
-0.5	&-	&-	&1.7717		&-	&-	&c) infinite \\
0		&-	&-	&2.0356		&-	&-	&d) infinite (dotted) \\
1		&-	&-	&2.4320		&-	&-	&e) infinite \\
\\
\hline
\end{tabular}
\end{table}

There is another interesting analysis devoted to finding the value of $d_*$ corresponding to the last stable circular orbit. Using the equations (\ref{eq21}) and (\ref{eq19}) we obtain
\begin{equation}
2u^2-3u-\varepsilon=0\ ,
\label{eq22}
\end{equation}
which takes into account both the conditions $\Phi(u)=0$ and $\Phi'(u)=0$. Equation (\ref{eq22}) yields two solutions corresponding to the critical point of the circular orbit $u_{circ}$ and the point of disappearance of closed trajectories $u_{lim}$
\begin{equation}
u_{circ}=\frac{3+\sqrt{9+8\varepsilon_{circ}}}{4}\ ,\ \ \ \ u_{lim}=\frac{3-\sqrt{9+8\varepsilon_{lim}}}{4}\ ,
\label{eq23}
\end{equation}
where from (\ref{eq21}), 
\begin{equation}
d_{circ}=u_{circ}^3 - u_{circ}^4 \ ,\ \ \ \ d_{lim}=u_{lim}^3 - u_{lim}^4 \ .
\label{eq23a}
\end{equation}

\section{Limiting parameters}
The two-body motion in the Kepler problem is characterized by a circular orbit at $\varepsilon=-1$, and elliptical orbits, with a large axis tending to infinity at $\varepsilon \rightarrow 0$ (see Fig.~\ref{fig1} and Table\,I). In the presence of DE, the circular orbits exist only in the limiting interval of the parameter $d$ values \cite{bkm2019}, and the transition from a finite to infinite trajectories happens abruptly, at a finite value of maximal separation. 

\subsection{Circular orbits}
\begin{figure}[t]
\includegraphics[width=1.0\textwidth]{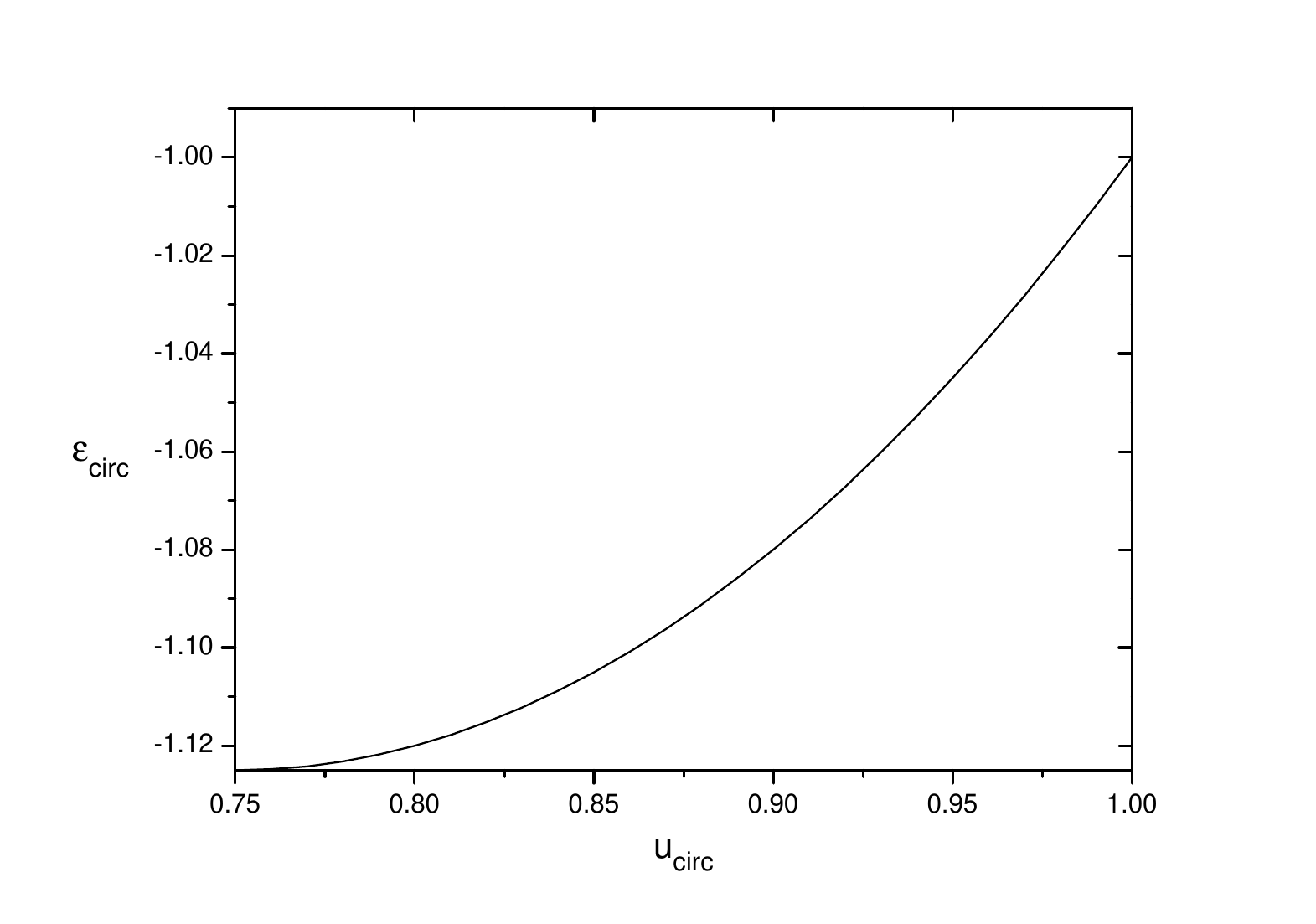}
\caption{The dependence of the dimensionless total energy 
$\varepsilon_{circ}$ on the dimensionless inverse radius $u_{circ}$ for circular orbits.
\label{fig6}}
\end{figure}

\begin{figure}[t]
\includegraphics[width=1.0\textwidth]{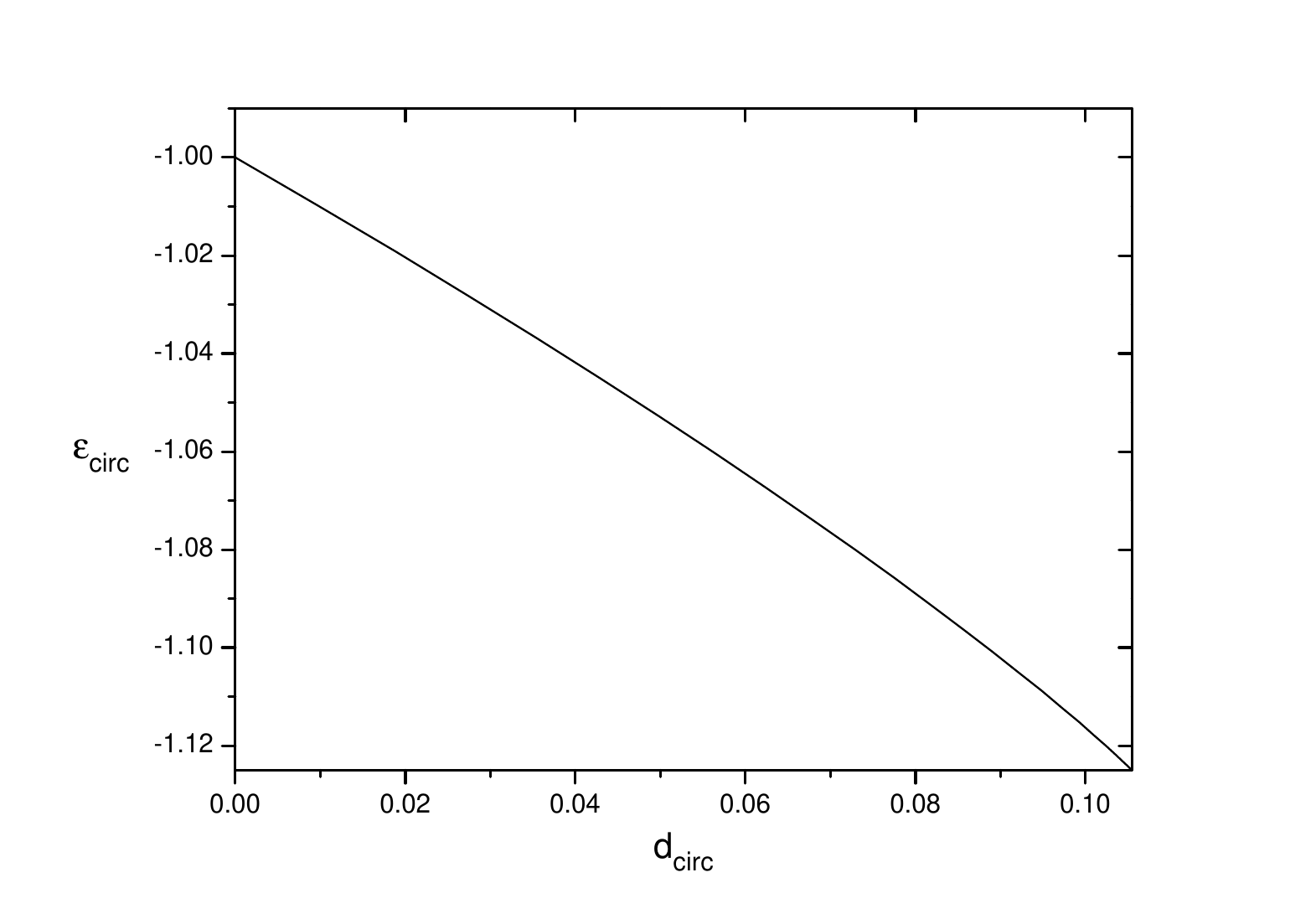}
\caption{The dependence of the dimensionless total energy
$\varepsilon_{circ}$ on the parameter $d_{circ}$ for circular orbits.
\label{fig7}}
\end{figure}

Consider first circular orbits, which were analyzed in detail in Ref.~\refcite{bkm2019}. It follows from the first of (\ref{eq23}) that circular orbits in the presence of DE exist only in the interval
\begin{equation}
-\frac{9}{8} < \varepsilon_{circ} < -1.
 \label{eq26}
\end{equation}
Here the left inequality follows from the need for a positive value inside the square root, and the right one is connected with a positive value of $d_{circ}$. The values $\varepsilon = -1$, $u_{circ}=1$, $d_{circ}=0$ correspond to the Keplerian motion in the absence of DE. The values
$$\varepsilon = -\frac{9}{8}, \quad u=\frac{3}{4}, \quad d=\frac{27}{256}\approx 0.1055$$
correspond to the maximum value of $d$, generating a cusp as previously discussed. Comparing this result with the corresponding one in Ref.~\refcite{bkm2019} where this extremum is characterized by the value $b_{lim}=r_k^3/2 r_0^3\approx 0.053$, we see that it agrees with our result since $d_{max}=2b_{lim}\approx 0.1055$. Note also that the stable part on the right plot in Fig.~3 of Ref.~\refcite{bkm2019}, obtained numerically, is now reproduced by the analytic relations (\ref{eq23}) and (\ref{eq23a}).
The dependence of $\varepsilon_{circ}$ on $u_{circ}$ and $d_{circ}$ is plotted in Figs.~\ref{fig6} and \ref{fig7}.

\begin{figure}[t]
\includegraphics[width=1.0\textwidth]{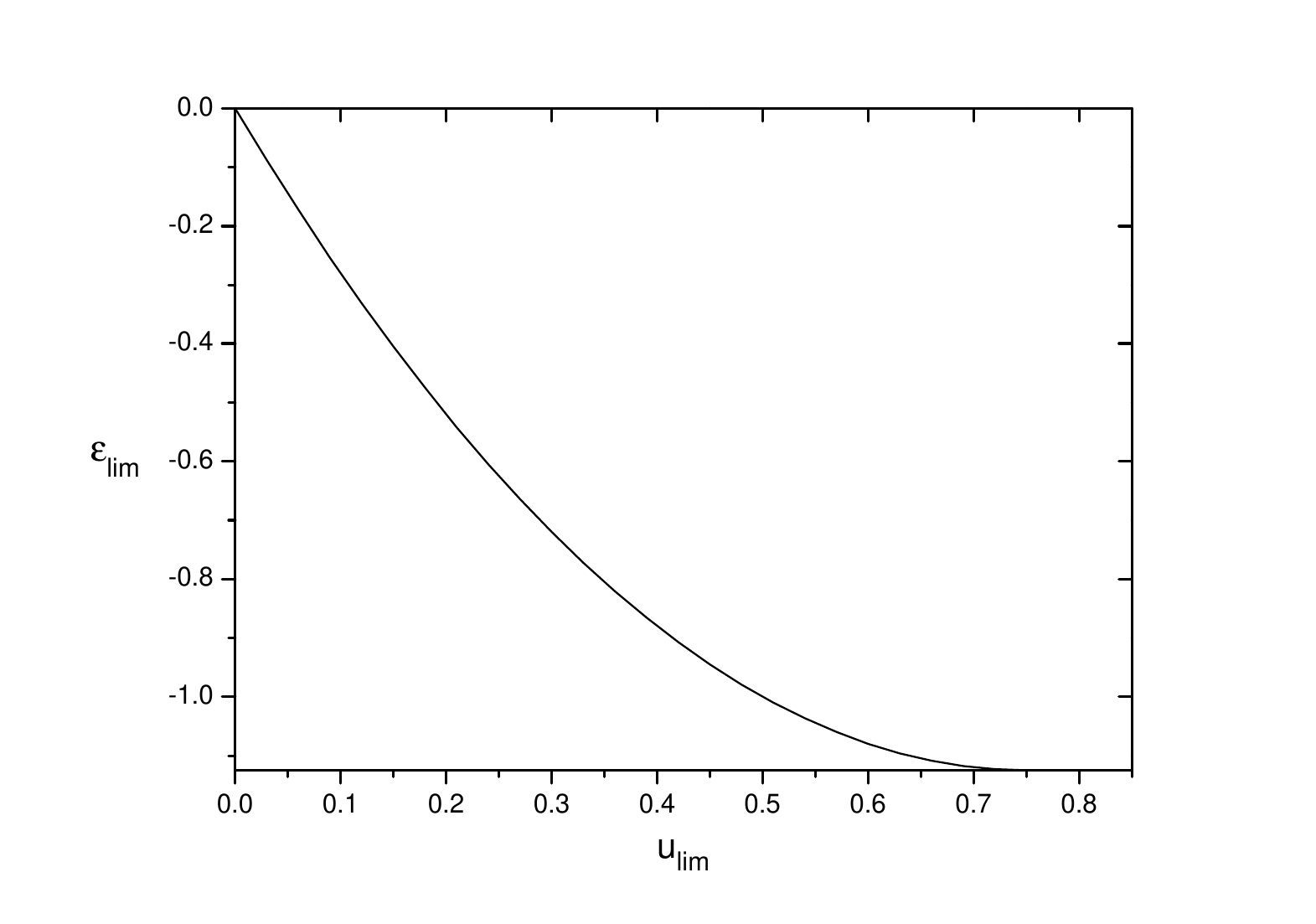}
\caption{The dependence of the dimensionless total energy
$\varepsilon_{lim}$ on the dimensionless inverse radius $u_{lim}$ for transition orbits.
\label{fig8}}
\end{figure}

\begin{figure}[t]
\includegraphics[width=1.0\textwidth]{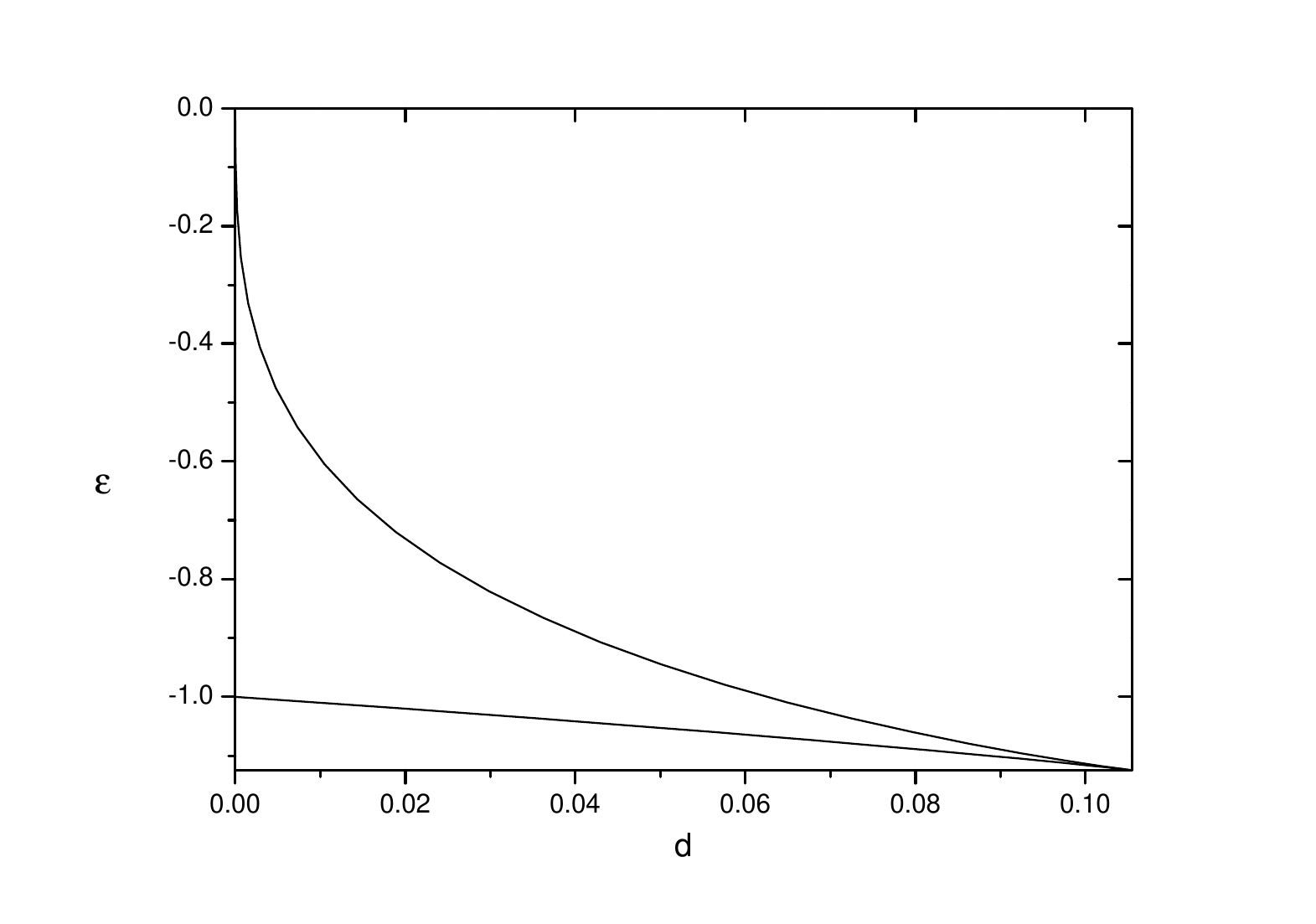}
\caption{The dependence of $\varepsilon_{lim}$ (upper curve) and 
$\varepsilon_{circ}$ (lower curve) on the parameter $d$. The lower curve is the same as in Fig.~\ref{fig6}. The finite orbits of the two-body motion in the presence of DE correspond to values of the parameters ($\varepsilon,d$) lying between these two curves.
\label{fig9}}
\end{figure}

\subsection{Limiting finite orbits}
The limiting orbits, separating finite and infinite motion in the presence of DE (transition orbits), exist in the following interval
\begin{equation}
-\frac{9}{8} \leq \varepsilon_{lim} < 0\ .
\label{eq27}
\end{equation}
Here the left inequality follows from the need for a nonnegative value inside the square root, and the right one is connected with the need for a positive value of $u_{lim}$ due to (\ref{eq23}). The values $\varepsilon = 0$, $u_{lim}=0$, $d_{lim}=0$ correspond to the parabolic orbit motion in the absence of DE. The limiting parameters for closed trajectories of the two-body motion in the presence of DE are presented in Figs.~\ref{fig8} and \ref{fig9}, plotted using the solutions (\ref{eq23}). Such transition orbits are shown in Figs.~\ref{fig2}-\ref{fig4} and the related parameters in Tables II-IV. Therefore, the limiting values for the parameters $u_{lim}$ and $d_{lim}$ are given by
\begin{equation}
\frac{3}{4}\geq u_{lim} \geq 0,\quad 27/256 > d_{lim} \geq 0, \quad {\rm for} \quad
-\frac{9}{8} \leq \varepsilon < 0\,.
\end{equation}

\section{Orbital precession in the presence of DE}

In the presence of DE the orbits are not closed (excluding circular motion in some cases)\cite{2013Emelyanov}, but it is convenient to consider a quasi-period for such motion, defined as the angular distance between two subsequent pericenters or apocenters of the trajectory. While the orbit is closed in the purely Keplerian case without DE, the change of this angular distance relative to $2\pi$ may be interpreted as an orbital pericenter precession due to the DE. The angular distance between two subsequent apocenters of the trajectory $\phi_{tb}$ using (\ref{eq11}) is
\begin{equation}
\phi_{tb} =2\int_{u_-}^{u_+}\frac{du}{\sqrt{-u^2 +2u +\varepsilon +d/u^2}},
\label{eq19a}
\end{equation}
where $u_{\pm}$ are the largest positive roots of the equation $-u^2+2u+\varepsilon +d/u^2=0$, with $u_- < u_+$.

\subsection{The case $\varepsilon=-1$}

For $\varepsilon=-1$ the roots can be found analytically
\begin{equation}
\begin{gathered}
u_0=\frac{1}{2}-\sqrt{\frac{1}{4}-\sqrt{d}},\quad
u_-=\frac{1}{2}+\sqrt{\frac{1}{4}-\sqrt{d}} \\
u_+=\frac{1}{2}+\sqrt{\frac{1}{4}+\sqrt{d}},\quad 
u_4=\frac{1}{2}-\sqrt{\frac{1}{4}+\sqrt{d}}\ .
\end{gathered}
\label{eq20a}
\end{equation}
Only the roots $u_+$ and $u_-$ define finite trajectories; $u_4$ is always negative and physically not relevant; $u_0$ can only define infinite trajectories. Numerical integration of (\ref{eq19a}) gives the results presented in Table VI. The precession angle 
$\phi_{pr}$, by definition equals
\begin{equation}
\phi_{pr} =\phi_{tb}-2\pi.
\label{eq21a}
\end{equation}
Therefore, the orbital precession in the presence of DE is the most important feature.
\begin{table}[h]
{{\bf Table VI.} Half angular distance 0.5\,$\phi_{tb}$ between two subsequent apocenters ($\varepsilon=-1$)}
\medskip
\centering
\scriptsize
\begin{tabular}{cccccc}
\hline
$d$   &$u_-$    &$u_+$     &$0.5\,\phi_{tb}$ (rad) &$\phi_{pr}$ (deg/cycle) &trajectory type \\
\hline
\\
0            & 1       & 1       & $\pi$    & 0       & circular \\
0.0001       &0.98990  &1.0099   &3.142064  &0.05404  &quasi-elliptical \\
0.001        &0.96731  &1.0307   &3.146347  &0.54478  &quasi-elliptical \\
0.002        &0.95308  &1.0429   &3.151186  &1.09933  &quasi-elliptical \\
0.005        &0.92343  &1.0663   &3.166246  &2.82504  &quasi-elliptical \\
0.01         &0.88730  &1,0916   &3.193357  &5.93176  &quasi-elliptical \\
0.02         &0.82951  &1.1256   &3.257160  &13.2431  &quasi-elliptical \\
0.03         &0.77712  &1.1505   &3.339735  &22.7054  &quasi-elliptical \\
0.04         &0.72361  &1.1708   &3.455523  &35.9737  &quasi-elliptical \\
0.05         &0.66246  &1.1882   &3.645932  &57.7930  &quasi-elliptical \\
0.06         &0.57107  &1.2035   &4.190171  &120.158  &quasi-elliptical \\
1/16         &0.50000  &1.2071   &$\infty$  & boundary of finite trajectories  & infinite \\
\\
\hline
\end{tabular}
\end{table}
As follows from the solutions (\ref{eq20a}), finite trajectories exist only when $d<1/16$ for 
$\varepsilon=-1$. At larger $d$ there is only one real positive root ($u_+$) defining an infinite trajectory. It is clear that finite motion in the presence of DE is possible only inside the zero gravity radius $r_0$ defined by the first equation of (\ref{eq4}). The presence of the centrifugal force, at finite angular momentum $L$, decreases the limiting value of the radius of the finite trajectory, so that actually $r_{lim} < r_0$. The dimensionless value of the zero gravity radius $x_0=r_0/r_k$ is directly defined by $d$, according to the definition of the parameter $d=(r_k/r_0)^3$:
\begin{equation}
x_0=d^{-1/3}.
 \label{eq22b}
\end{equation}
For $\varepsilon=-1$ we have $x_0=(1/16)^{-1/3}\approx 2.52$. As follows from the Table VI, the value of the limiting radius of finite trajectories corresponds to $u_-$ at $d=1/16$, namely $u_{lim}=u_-=0.5$. This means that $x_{lim}=1/u_{lim}=2<x_0\approx 2.52$.

\subsection{Periods in the quasi-periodic two-body motion in the presence of DE}

The dimensionless quasi-period ${\tilde P_{tb}}$ of the two-body motion in the presence of DE is
\begin{equation}
{\tilde P_{tb}}=2\int_{u_-}^{u_+}\frac{du}{u^2\sqrt{-u^2+2u+\varepsilon+d/u^2}}.
\label{eq42}
\end{equation}
Then the quasi-period $P_{tb}$ in units of the Keplerian period $P_k$ at the same $\varepsilon$ and $d=0$ is given by
\begin{equation}
\frac{P_{tb}}{P_k}={\tilde P_{tb}}\,\frac{(-\varepsilon)^{3/2}}{2\pi}\ .
\label{eq42a}
\end{equation}
In the Kepler problem the period is a function of one parameter $\varepsilon$ (see Eq.~(\ref{eq16})), while in the presence of DE the quasi-period also depends on $d$.

\section{Trajectories at $L=0$}
As mentioned above, the limiting value of the radius for finite trajectories is less than the zero-gravity radius $r_0$, due to additional repulsion from a centrifugal force. Only in the case of zero angular momentum does the limiting radius coincide with the zero-gravity radius. The previous dimensionless considerations cannot be applied to the case with $L=0$ because the scaling radius $r_k$ vanishes. It is easy to analyze this case in the original dimensional variables. From equations (\ref{eq2}) and (\ref{eq3}) we have
\begin{equation}
E=\frac{1}{2}\mu\dot{r}^2-\mu\left(\frac{GM}{r}+\frac{\Lambda c^2}{6}r^2\right),\quad \dot\varphi=0.
\label{eq29}
\end{equation}
Analogous to (\ref{eq5}) the equation for radial dependence on time is
\begin{equation}
 \frac{dr}{dt} =\pm \sqrt{2\frac{E}{\mu}+2\left(\frac{GM}{r}+\frac{\Lambda c^2}{6}r^2\right)},\quad \Psi(r)=2\frac{E}{\mu}+2\left(\frac{GM}{r}+\frac{\Lambda c^2}{6}r^2\right).
\label{eq30}
\end{equation}
The zeros of the function $\Psi(r)$ define the turning points of the linear trajectory,
with nonlinear oscillations.\footnote{Here Newtonian theory is used at all velocities, which formally become superluminal at small $r$. We are interested here, mainly in the boundary between finite and infinite trajectories, which are found correctly in this approximation.} The boundary between finite and infinite trajectories is a saddle point of the trajectory where $\Psi$ and its derivative become zero
\begin{equation}
\Psi(r)=2\frac{E}{\mu}+2\left(\frac{GM}{r}+\frac{\Lambda c^2}{6}r^2\right)=0, \quad \Psi^{'}= -\frac{2GM}{r^2}+\frac{2\Lambda c^2}{3}r=0.
\label{eq31}
\end{equation}
This system determines the boundary between finite and infinite trajectories $r_{lim}$, and value of the energy $E_{lim}$, at which this boundary is reached (see Eq.~(\ref{eq4}))
\begin{equation}
r_{lim}=\left(\frac{3GM}{\Lambda c^2}\right)^{1/3}=r_0, \quad E_{lim}=-\frac{3}{2}\frac{\mu GM}{r_0}.
\label{eq32}
\end{equation}
The finite amplitude oscillations occur only for $E<E_{lim}$; at larger $E$ the trajectory goes to infinity. To solve the equation for the linear trajectory, we use the first equation of (\ref{eq30}) in dimensionless variables, introducing an arbitrary radius $r_*$
\begin{equation}
x=\frac{r}{r_*}, \quad \tau=\frac{t}{t_*}, \quad t_*=\frac{r_*^{3/2}}{\sqrt{GM}},\quad \varepsilon=\frac{2Er_*}{\mu GM}, \quad \frac{\Lambda c^2 r_*^3}{6GM}=\frac{1}{2}\left(\frac{r_*}{r_0}\right)^3 ,
\label{eq33}
\end{equation}
in the form
\begin{equation}
\frac{d\tau}{dx}=\pm \frac{1}{\sqrt{(r_*/r_0)^3 x^2 +2/x +\varepsilon}}.
\label{eq34}
\end{equation}
It is convenient to use the variable $y=1/x$, and Eq.~(\ref{eq34}) assumes the form
\begin{equation}
\frac{d\tau}{dy}=\mp\frac{1}{y^2\sqrt{(r_*/r_0)^3 /y^2 +2y +\varepsilon}}.
\label{eq35}
\end{equation}
Here $\varepsilon \leq -3\,(r_*/r_0)$, according to (\ref{eq32}) and (\ref{eq33}).  At $\Lambda>0$ we may use $r_*=r_0<\infty$, and Eq.~(\ref{eq35}) takes the simpler form
\begin{equation}
\frac{d\tau}{dy}=\mp\frac{1}{y^2 \sqrt{\varepsilon+2y+1/y^2}}.
\label{eq36}
\end{equation}

In the presence of a singularity at $r=0$, there are two possible interpretations of the oscillations. They may be interpreted as a limiting trajectory of two-body motion as $L\rightarrow 0$. At any nonzero $L$ the singularity is avoided, and the period of oscillations is defined as
\begin{equation}
P_{tb0}=2\int_{y_-}^{\infty}\frac{dy}{y^2\sqrt{\varepsilon+2y+1/y^2}}.
\label{eq37}
\end{equation}
In the second interpretation the two-body motion with zero angular momentum is passing through the singularity and Eq.~(\ref{eq36}) describes the oscillating motion between the points $y=+y_-$ and $y=-y_-$, where two bodies cross through each other and exchange their positions. In this case the period of oscillations is equal to $2P_{tb0}$.
\\
In absence of DE we have from Ref.~\refcite{ryzhik}, and equations (\ref{eq35}) and (\ref{eq37}), the expression for the limiting Keplerian period $P_{k0}$ at $L=0$
\begin{eqnarray}
\frac{d\tau}{dy}=\mp\frac{1}{y^2 \sqrt{\varepsilon +2y}}, \quad
P_{k0}=2\int_{-\varepsilon/2}^{\infty}\frac{dy}{y^2\sqrt{\varepsilon+2y}}= \nonumber \\
2\left[{\frac{\sqrt{\varepsilon+2y}}{-\varepsilon y}+\frac{2}{(-\varepsilon)^{3/2}}\arctan\sqrt{\frac{\varepsilon+2y}{-\varepsilon}}}\right]_{-\varepsilon/2}^{\infty}=\frac{2\pi}{(-\varepsilon)^{3/2}}.
 \label{eq38}
\end{eqnarray}
Comparing with (\ref{eq35}), we see that the Keplerian period oscillations is defined by the same expression at all $L\ge 0$. Taking into account the result (\ref{eq38}), we can express the period of linear oscillations (\ref{eq37}) at $L=0$ in units of the Keplerian period $\tilde P_{tb0}$ as
\begin{equation}
{\tilde P_{tb0}}= P_{tb0}\frac{(-\varepsilon)^{3/2}}{2\pi}.
\label{eq39}
\end{equation}

\section{Conclusions}
We considered the Keplerian two-body problem with non-circular orbits, in the presence of dark nergy (identified with the cosmological constant $\Lambda$) introduced as a third additional force. The values of dimensionless parameters determining the typology of trajectories  for variable $\Lambda$ (or equivalently $d$) and $\epsilon$ are determined. It is found that in the presence of a dark energy only two types of trajectories are present.
  
1. Pure unbound trajectories for a family of parameters, corresponding to very large distance between the two gravitating bodies at a large negative total energy of the pair, or parameters corresponding to positive total energy. 

2. Simultaneous existence of bounded and unbounded trajectories for the same combination of parameters, defining the intermediate values of the total energy.
 
The bound trajectories could be represented approximately by precessing quasi-elliptical trajectories for small values of 
$\Lambda$, treated as small perturbation of the Keplerian elliptical orbit. Critical parameter values are found which determine the points at which the spontaneous transition from bound to unbound orbits can occur.

\section*{{\centerline{Appendix}}}
{\bf{\centerline {Orbit precession, and binary  period correction, in a linear}}}
{\bf{\centerline {approximation, for small values of $\Lambda$}}}
\medskip
A study of the orbital precession in linear approximation was studied earlier in Refs.~\refcite{neisht,khm03}, using rather complicated methods of time averaging of equations following from Lagrangian and Hamiltonian functions. Here we obtain the precession frequency, and corrections to the Keplerian period, for small influence of $\Lambda$ (small values of $d$), in linear approximation for small valued of $d$, using a simple method considered in Ref.~\refcite{1969Landau}, which permitted to avoid non-physical singularities.

We find the precession frequency $\omega_{pr}$, calculating precession angle, in linear approximation, during one Keplerian period \eqref{eq15}, using equations \eqref{eq19a} and \eqref{eq21a}.
\begin{equation}
\begin{gathered}
\omega_{pr}=\frac{\phi_{pr}}{P_k}, \qquad \phi_{pr}=\phi_{tb}-2\pi, \\
\phi_{tb} =2\int_{u_-}^{u_+}\frac{du}{\sqrt{-u^2+2u+\varepsilon +d/u^2}}=4\,\frac{\partial}{\partial\varepsilon}\bigg[\int_{u_-}^{u_+}\sqrt{-u^2+2u+\varepsilon+d/u^2}\ du \bigg]\\
=2\int_{u_-}^{u_+}\frac{du}{\sqrt{-u^2+2u+\varepsilon}}+
2d\,\frac{\partial}{\partial\varepsilon}\bigg[\int_{u_-}^{u_+}\frac{du}{u^2\sqrt{-u^2+2u+\varepsilon}}\bigg]\\
=2\pi+2d\,\frac{\partial}{\partial\varepsilon}\bigg[\int_{0}^{\pi}\frac{d\varphi}{u^2(\varphi)}\bigg]=2\pi+2d\,\frac{\partial}{\partial\varepsilon}\bigg[\int_{0}^{\pi}\frac{d\varphi}{(1-e\cos\varphi)^2}\bigg].
\end{gathered}
\label{eqa1}
\end{equation}
Here it is taken into account Eq.~\eqref{eq18}, from which
\begin{equation}
u(\varphi)=1+e\sin(\varphi-\pi/2)=1-e\cos\varphi,\quad e=\sqrt{1+\varepsilon}, \quad d\varphi =\frac{du}{\sqrt{-u^2+2u+\varepsilon}}
\label{eqa2}
\end{equation}
The last integral in \eqref{eqa1} is present in Ref.~\refcite{ryzhik}, from where we have
\begin{equation}
\int_{0}^{\pi}\frac{d\varphi}{(1-e\cos\varphi)^2}=\frac{1}{1-e^2}\int_{0}^{\pi}\frac{d\varphi}{1-e\cos\varphi}=\frac{\pi}{(1-e^2)^{3/2}}=\frac{\pi}{(-\varepsilon)^{3/2}}.
\label{eqa3}
\end{equation}
From these equations, with parameters of the Kepler motion obtained by equations \eqref{eq16} and \eqref{eq4},
\begin{equation}
P_k=2\pi \frac{a^{3/2}}{\sqrt{GM}}, \quad a=\frac{L^2}{\mu^2 GM(1-e^2)}, \quad\omega_k=\frac{2\pi}{P_k}=\frac{\sqrt{GM}}{a^{3/2}}, \quad t_0=\frac{\mu r_k^2}{L},
\label{eqa4}
\end{equation}
we get the precession angle during one Keplerian period $\phi_{pr}$, and precession frequency $\omega_{pr}$ in the form in which they had been derived in Refs.~\refcite{neisht,khm03}:
\begin{equation}
\begin{gathered}
\phi_{pr}=\frac{3\pi d}{(-\varepsilon)^{5/2}}, \quad d=\left(\frac{r_k}{r_0}\right)^3 =\frac{\Lambda c^2 L^6}{3\mu^6(GM)^4},\\
\phi_{pr}=\frac{3\pi d}{(1-e^2)^{5/2}}=\frac{\pi\Lambda c^2 L^6}{\mu^6(GM)^4(1-e^2)^{5/2}}=\frac{\pi\Lambda c^2 a^3\sqrt{1-e^2}}{GM},\\
\omega_{pr}=\frac{\phi_{pr}}{P_k}=\frac{\Lambda c^2 \sqrt{1-e^2}}{2\omega_k}
=\frac{\Lambda c^2 a^{3/2}\sqrt{1-e^2}}{2\sqrt{GM}}=\frac{3d}{-2\varepsilon t_0}=\frac{3d}{2(1-e^2)t_0}.
\end{gathered}
\label{eqa5}
\end{equation}

To find corrections to the period between two subsequent apocenter transitions of the given point of the trajectory, we start from \eqref{eq42}. We have
\begin{equation}
\begin{gathered}
{\tilde P}_{tb}=2\int_{u_-}^{u_+}\frac{du}{u^2\sqrt{-u^2+2u+\varepsilon +d/u^2}} =4\,\frac{\partial}{\partial\varepsilon}\bigg[\int_{u_-}^{u_+}\frac{du}{u^2}\sqrt{-u^2+2u+\varepsilon +d/u^2}\bigg]\\
=2\int_{u_-}^{u_+}\frac{du}{u^2\sqrt{-u^2+2u+\varepsilon}}+
2d\,\frac{\partial}{\partial\varepsilon}\bigg[\int_{u_-}^{u_+}\frac{du}{u^4\sqrt{-u^2+2u+\varepsilon}}\bigg]\\
=\frac{2\pi}{(-\varepsilon)^{3/2}}+2d\,\frac{\partial}{\partial\varepsilon}\bigg[\int_{0}^{\pi}\frac{d\varphi}{u^4(\varphi)}\bigg]
=\frac{2\pi}{(-\varepsilon)^{3/2}}+2d\,\frac{\partial}{\partial\varepsilon}\bigg[\int_{0}^{\pi}\frac{d\varphi}{(1-e\cos\varphi)^4}\bigg].
\end{gathered}
\label{eqa6}
\end{equation}
For derivation of analytic expression for the binary period in the presence of DE, as a perturbation, we need to calculate the following integrals, using Ref.~\refcite{ryzhik}
\begin{equation}
\begin{gathered}
\int_{0}^{\pi}\frac{d\varphi}{(1-e\cos\varphi)^4}=
\frac{1}{1-e^2}\int_{0}^{\pi}\frac{d\varphi}{(1-e\cos\varphi)^3}+\frac{2e}
{3(1-e^2)}\int_{0}^{\pi}\frac{\cos\varphi\,d\varphi}{(1-e\cos\varphi)^3},\\
\int_{0}^{\pi}\frac{d\varphi}{(1-e\cos\varphi)^3}=
\frac{1}{1-e^2}\int_{0}^{\pi}\frac{d\varphi}{(1-e\cos\varphi)^2}+\frac{e}
{2(1-e^2)}\int_{0}^{\pi}\frac{\cos\varphi\,d\varphi}{(1-e\cos\varphi)^2}.
\end{gathered}
\label{eqa7}
\end{equation}
\begin{equation}
\begin{gathered}
\int_{0}^{\pi}\frac{\cos\varphi\,d\varphi}{(1-e\cos\varphi)^3}=\frac{1}
{2(1-e^2)}\int_{0}^{\pi}\frac{(2e+\cos\varphi)\,d\varphi}{(1-e\cos\varphi)^2},\\
\int_{0}^{\pi}\frac{\cos\varphi\,d\varphi}{(1-e\cos\varphi)^2}=\frac{e}{1-e^2}\int_{0}^{\pi}\frac{d\varphi}{1-e\cos\varphi}=\frac{\pi e}{(1-e^2)^{3/2}}.
\end{gathered}
\label{eqa8}
\end{equation}
We have from equations \eqref{eqa3} and \eqref{eqa6}-\eqref{eqa8}
\begin{equation}
\begin{gathered}
I_4=\int_{0}^{\pi}\frac{d\varphi}{(1-e\cos\varphi)^4}=\frac{\pi}{(1-e^2)^{7/2}}\bigg(1+\frac{3}{2}e^2\bigg)=\frac{\pi}{(-\varepsilon)^{7/2}}\bigg(\frac{5}{2}+\frac{3}{2}\varepsilon \bigg),
\end{gathered}
\label{eqa9}
\end{equation}
\begin{equation}
\begin{gathered}
{\tilde P}_{tb}=\frac{2\pi}{(-\varepsilon)^{3/2}}+2d\,\frac{\partial I_4}
{\partial\varepsilon},\\
\frac{\partial I_4}{\partial\varepsilon}=\frac{5\pi}{(1-e^2)^{9/2}}\bigg[\frac{7}{4}-\frac{3}{4}(1-e^2)\bigg]
=\frac{5}{4}\frac{\pi}{(1-e^2)^{9/2}}(4+3e^2),\\
\Delta P_{pr}= P_{tb}-P_k=(\tilde P_{tb}-\tilde P_k)\,t_0= 2dt_0\frac{\partial I_4}{\partial\varepsilon}=\frac{5\pi}{6}\frac{\Lambda c^2}{\omega_k^3} (4+3e^2).
\end{gathered}
\label{eqa10}
\end{equation}

\end{document}